# Inhibition of light tunneling in waveguide arrays

A. Szameit[1], Y. V. Kartashov[2], F. Dreisow[1], M. Heinrich[1], T. Pertsch[1], S. Nolte[1], A. Tünnermann[1], V. A. Vysloukh[3], F. Lederer[4], and L. Torner[2]

[1]Institute of Applied Physics, Friedrich Schiller University Jena, Max-Wien-Platz 1, 07743 Jena, Germany

[2]ICFO-Institut de Ciencies Fotoniques, and Universitat Politecnica de Catalunya, Mediterranean Technology Park, 08860 Castelldefels (Barcelona), Spain

[3]Departamento de Fisica y Matematicas, Universidad de las Americas – Puebla, 72820, Puebla, Mexico

[4]Institute for Condensed Matter Theory and Optics, Friedrich-Schiller-University Jena, Max-Wien-Platz 1, 07743 Jena, Germany

We report the observation of almost perfect light tunneling inhibition at the edge and inside laser-written waveguide arrays due to band collapse. When the refractive index of the guiding channels is harmonically modulated along the propagation direction and out-of-phase in adjacent guides, light is trapped in the excited waveguide over a long distance due to resonances. The phenomenon can be used for tuning the localization threshold power.

The precise control of wave packet tunneling by external driving fields is of major relevance in many branches of physics, such as superconducting quantum interference devices, spin systems, multiquantum dots and cold atoms in optical traps [1,2]. Two different phenomena attract particular interest: Dynamic localization in longitudinally periodic potentials [3-5] and driven double-well potentials [6,7] which are well suited to investigate tunneling control. Optical settings provide a new system to explore tunneling phenomena [8,9] as well as diffraction-free wave packet propagation [10,11]. On this regard, a particularly important system is put forward by arrays of evanescently coupled waveguides, where it was shown that the concept of inhibited light tunneling is not only possible via lattice soliton formation [12-15], but also due to a harmonic bending of the waveguides yielding ei-ther dynamic localization (DL) [16-20] or coherent destruction of tunneling (CDT) [21,22]. However, while DL occurs only in systems without boundary interaction, CDT was achieved only in a two-waveguide system due to analogies with surface states in curved lattices [23].

In this Letter we demonstrate that a harmonic out-of-phase modulation of the linear refractive index along the propagation direction yields the almost perfect inhibition of the light tunneling between adjacent guiding channels irrespective of the input position in finite and infinite arrays. When the frequency and amplitude of the modulation are properly chosen the band of quasi-energies is considerably narrowed, forcing the light to remain in the excited channel. This phenomenon is possible in the coupler geometry, at the edge and in the interior of waveguide arrays. At intermediate power levels, the light partially delocalizes and eventually relocalizes again due to soliton formation at high power levels.

To gain intuitive insight, we start our analysis by studying the dimensionless equations describing propagation of light in the waveguide array in tight-binding approximation

$$i\frac{dq_m}{d\xi} + (-1)^m \mu \sin(\Omega\xi) q_m + C(q_{m+1} + q_{m-1}) + \chi q_m |q_m|^2 = 0. \tag{1}$$

which describes the evolution of the amplitude in the $m$-th waveguide $q_m$, with light tunneling into adjacent guides with the tunneling rate $C$ and the nonlinearity constant $\chi$. The value $1 > \mu \geq 0$ is the relative depth of the harmonic longitudinal modulation, while $\Omega$ is its spatial frequency. The modulation of the refractive index between the adjacent guiding channels is out-of-phase. The transformation $h_m = q_m \exp[i(-1)^m \mu \cos\{\Omega\xi\}/\Omega]$ yields

$$i\frac{dh_m}{d\xi} + C(h_{m+1} + h_{m-1})\exp[2i\mu\cos(\Omega\xi)/\Omega] + \chi h_m |h_m|^2 = 0 \tag{2}$$

When using the expansion $\exp[2i\mu\cos(\Omega\xi)/\Omega] = \sum_k i^k J_k(2\mu/\Omega)\exp(ik\Omega\xi)$ in terms of Bessel functions and neglecting all orders except $k = 0$, one finds that diffraction vanishes when $(2\mu/\Omega) = \nu_j$ with $\nu_j \approx 2.4, 5.5, ...$ being roots of the zero-order Bessel function. Hence, for a fixed modulation depth $\mu$, such crude approximation predicts that there are resonance frequencies at which light tunneling is inhibited. In the limiting case of a linear optical coupler ($\chi \to 0$) when only the first channel is exited with unit amplitude, the solution of Eq.

(2) is $|h_1(\xi)|^2 = [1 + \cos(2\kappa\xi)]/2$, where the coupling constant $\kappa = CJ_0(2\mu/\Omega)$ is reduced by the factor $J_0(2\mu/\Omega)$. The distance-averaged power fraction guided in the excited channel $U_m = L^{-1}\int_0^L |h_1(\xi)|^2 d\xi$, can be found analytically as $U_m = \{1 + \text{sinc}[2CLJ_0(2\mu/\Omega)]\}/2$. Thus, around the zeroes of the Bessel function, the power fraction can be estimated as $U_m \simeq 1 - C^2L^2J_0^2(2\mu/\Omega)/3$. When $J_0(2\mu/\Omega) \to 0$ the localization is complete. In the nonlinear case, tunneling inhibition implies a decreasing of the critical power $P_{cr} = 4CJ_0(2\mu/\Omega)/\chi$, which is proportional to the reduction of the coupling constant. In the under-critical nonlinear case the period of the power oscillations is defined by an elliptic integral of the first kind, it grows monotonically and diverges as the input power $P_0$ approaches the critical one. The corresponding solution of Eq. (2) is $|h_1(\xi)|^2 = [1 + \text{cn}(2\kappa\xi, k)]/2$, where $k = P_0^2/P_{cr}^2$. In the general case, localization is described by the Floquet-Bloch formalism, in which every excitation is a superposition of discrete Bloch waves [24]. The entire set of Bloch waves results in the formation of a quasi-energy bands, yielding spatial dispersion and, therefore, light tunneling into adjacent guides. At the resonance condition the bands flatten, preventing light from spreading into the array. Tunneling inhibition cannot be exact [24].

To elucidate a more rigorous dynamics we conducted simulations with the nonlinear Schrödinger equation for the dimensionless field amplitude $q$, which describes the propagation of light along the $\xi$-axis of waveguide array under the assumption of cw radiation:

$$i\frac{\partial q}{\partial \xi} = -\frac{1}{2}\frac{\partial^2 q}{\partial \eta^2} - |q|^2 q - pR(\eta,\xi)q. \qquad (3)$$

Here $\eta$ and $\xi$ are the transverse and longitudinal coordinates, while the parameter $p$ describes the refractive index modulation depth. The refractive index profile of the lattice is given by $R(\eta,\xi) = \sum_{m=-(M-1)/2}^{(M-1)/2} [1 + (-1)^m \mu \sin(\Omega\xi)]\exp[-(\eta - mw_s)^6/w_\eta^6]$. The super-Gaussian refractive index profile of the individual channels is fitted to the shape of the fabricated waveguides [25] and is characterized by the normalized width $w_\eta$. The parameter $w_s$ stands for the waveguide spacing, $M$ is the total number of the guiding channels. As the input condition we used $A\exp(-\eta^2/W^2)$ with $W$ being the beam width. For the fixed propagation distance $L$, the quality of the localization was characterized by the distance-averaged power fraction trapped in the excited waveguide channel
$U_m = L^{-1} \int_0^L d\xi \int_{-w_s/2}^{w_s/2} |q(\eta,\xi)|^2 d\eta \Big/ \int_{-w_s/2}^{w_s/2} |q(\eta,0)|^2 d\eta$.

Our samples were fabricated using a femtosecond-writing method (see [15] for full details of the fabrication method). In addition to the high flexibility of this method concerning the waveguide paths, a high degree of freedom is also possible for the refractive index distribution of the individual guides. Since the index modulation of the single lattice sites crucially depends on the writing velocity, one can particularly introduce an out-of-phase longitudinal harmonic modulation of the trapping channels by varying slightly the writing speed for each waveguide. Since the average writing velocity was $\sim 1750$ $\mu$m/s, the focusing nonlinearity is spatially uniform ($n_2 = 2.7 \times 10^{-20}$ m$^2$/W). In all our samples the width of the individual guides amounts to 3 $\mu$m, which is equivalent to $w_\eta = 0.3$. We fabricated two kinds of waveguide arrays: one having the spacing of 14 $\mu$m ($w_s = 1.4$) and the length $L = 40$ mm for low-power experiments at the visible wavelength $\lambda = 633$ nm and the other one with spacing 36 $\mu$m ($w_s = 3.6$) and length $L = 105$ mm for analyzing the high-power propagation of a fs laser beam of a Ti: Sapphire laser at $\lambda = 800$ nm. The first sample allows for the direct observation of the linear propagation inside our arrays using a special fluorescence technique [25], while in the second one the spacing of the lattice sites was increased so that the nonlinearity can overcome the evanescent coupling. Nevertheless, the excitation at a higher wavelength and the increased length of the guides in the latter sample yield a similar net diffraction in both arrays which ensures the comparability of the samples.

In a first step we demonstrate the light tunneling inhibition in a two-waveguide coupler. To evaluate the specific frequency of the longitudinal refractive index modulation, we fabricated a non-modulated optical coupler ($\mu = 0$), whose low power propagation pattern is shown in Fig. 1(a) at $\lambda = 633$ nm. The simulations shown in Fig. 1(b) yield the refractive index modulation depth $\approx 4 \times 10^{-4}$. The measured intensity beating frequency was $\Omega_b = 0.386$ mm$^{-1}$. On the basis of these values we calculated the optimal longitudinal modulation frequency of the refractive index yielding maximal $U_m$ in a coupler to be $\Omega_r/\Omega_b = 1.25$ at $\mu = 0.2$. Figure 1(c) shows the linear light tunneling inhibition in our samples in comparison with the corresponding modeling results [Fig. 1(d)].

When solving Eq. (3) one obtains that the frequency of the principal resonance grows monotonically with $\mu$ [Fig. 2(a)] and tends to $\Omega_b/2$ at $\mu \to 0$. Additionally, the dependence of $U_m$ on the modulation frequency is nontrivial, exhibiting several resonance peaks [Fig. 2(b)]. Both results are well fitted by the discrete model [Eq. (1)], in which the light tunneling inhibition in a two-waveguide geometry corresponds to the degeneracy point of the propagation constants of the $\xi$-periodic Floquet-Bloch states. In particular the well-defined resonance peak in Fig. 2(b) corresponds to the first root of $J_0(2\mu/\Omega) = 0$. Therefore, the main factor for tunneling inhibition is the periodic phase shift of the propagating modes introduced in Eq. (1). However, note that the additional factor appears only in the continuous model: The fundamental mode profile of a waveguide breathes following the $\xi$-oscillations of the waveguide depth as readily visible in Fig. 1(d), which is not included in the discrete model of Eq. (1). Accordingly, this results in more complicated resonance curve.

As mentioned above, nonlinearity slows down the power oscillations (see [26] for details). Therefore, if the modulation frequency is equal to or lower than the resonant one ($\Omega \leq \Omega_r$), an increase of the normalized peak intensity $A^2$ of the input beam shifts the power oscillations frequency away from the resonance peak thus leading to initial delocalization, while re-localization appears at higher powers due to soliton-type formation [see Fig. 2(c) where $\Omega$ matches the principal linear resonance]. For $\mu = 0.1$ when $\Omega_r = 0.76\Omega_b$, the minimal localization corresponds to different amplitudes $A^2 = 0.70$, $0.82$, $0.90$ for $\Omega/\Omega_b = 0.66$, $0.71$, $0.76$, respectively, but for all these frequencies relocalization occurs approximately for the same amplitude $A^2 \simeq 1.05$. In contrast, when $\Omega > \Omega_r$ nonlinearity shifts the frequency of power oscillations towards resonant values, thus producing localization enhancement from the very beginning. For example, at $\mu = 0.1$, $\Omega_r = 0.76\Omega_b$ the localization maximum appears at $A^2 = 0.46$ and $0.70$ for $\Omega/\Omega_b = 0.81$ and $0.86$, respectively. These amplitudes are smaller than the critical value $A^2 \simeq 1.4$ at which localization occurs in the unmodulated coupler ($\mu = 0$), a result that indicates that out-of-phase longitudinal modulation of refractive index might be used for fine tuning the localization threshold power.

When these results were analyzed for a waveguide array ($M = 13$, as in the experiment) we found out that the linear resonance curves are qualitatively similar for excitations of the edge channel and central channel [compare curves 1 and 2 in Fig. 2(d)]. However, the principal peak is more pronounced in the case of the edge channel excitation, because of the diminished discrete diffraction. Besides the most pronounced principal resonance at $\Omega_r \approx 1.30\Omega_b$, additional weaker peaks appear close to the fractional frequencies $\Omega_r/2, \Omega_r/3,\ldots$. The data of simulations were used for the optimization of the modulation frequency of fabricated arrays. We found numerically that for a modulation depth of $\mu = 0.2$ the optimal longitudinal frequency is $\Omega_r \approx 1.30\Omega_b$ for the surface channel excitation and $\Omega_r \approx 1.38\Omega_b$, when a waveguide in the array center is excited. Figure 3 compares the light propagation in non-modulated and optimally modulated waveguide arrays for the edge [panels (a),(b)] and central channel excitations [panels (c),(d)]. This is a generalization of the tunneling control in a double well potential. Due to the modulation of the refractive index the Floquet-Bloch modes exhibit almost identical quasi-energies, irrespective of the number of waveguides in the system or the position of excitation. Note that the possibility of linear light localization in the bulk or at the surface of arrays expands the opportunities for diffraction control and spatio-spectral selectivity of light localization.

To observe the impact of nonlinearity we monitored the power-dependent tunneling inhibition with femtosecond-pulsed radiation ($\tau_{\text{pulse}} = 150$ fs) of a Ti:Sa laser. In the experiments we used a longitudinal modulation depth of $\mu = 0.15$ and a longitudinal frequency of $\Omega_r \approx 1.05\Omega_b$ for surface excitation and $\Omega_r \approx 1.07\Omega_b$ for the excitation of the center waveguide. In Fig. 4 every subplot consists of a theoretical panel, showing the light intensity spatial distribution inside the sample, on top of the photograph of the experimentally observed output patterns. Columns (a),(c) show the transformation of light tunneling in a non-modulated array to a soliton-type tunneling inhibition with increasing input power; columns (b),(d) illustrate linear tunneling inhibition in a modulated array, partial delocalization at intermediate power level, and finally soliton formation at high input power. This behavior corresponds to the simulations of the power dependence of the localization parameter $U_m$. At low power, the resonant light propagation results in the inhibition of light diffraction. For an increased intermediate power, the nonlinear influence distorts tunneling inhibition, so that light can couple from the excited into the adjacent guides. However, at high input power, a soliton-like localization occurs due to the Kerr effect. It is interesting to note, that this is a representation of a diffraction-managed soliton, which were demonstrated experimentally only recently [27]. However, in our system it is possible to obtain soliton formation for both: increased power when the resonance condition is satisfied, and for de-creased power when the propagation is slightly off-resonant and the resonance curve is broadened by the nonlinear influence.

In conclusion, we observed experimentally light tunneling inhibition in waveguide arrays with a harmonic out-of-phase longitudinal modulation of the refractive index. The setup is a generalization of a double-well potential and allows full control of tunneling in an extended potential. These results indicate that resonant phenomena accessible in longitudinally modulated structures open new ways for the control of light propagation.

The authors wish to thank S. Longhi for very useful discussions.

# References

1. M. Grifoni et al., Phys. Rep. **304**, 229 (1997).
2. S. Kohler et al., Phys. Rep. **406**, 397 (2005).
3. D. H. Dunlap et al., Phys. Rev B **34**, 3625 (1986).
4. M. M. Digham et al., Phys. Rev. Lett. **88**, 046806 (2002).
5. H. Lignier at al., Phys. Rev. Lett. **99**, 220403 (2007).
6. F. Grossman et al., Phys. Rev. Lett. **67**, 516 (1991).
7. E. Kierig et al., Phys. Rev. Lett. **100**, 190405 (2008).
8. U. Peschel et al., Opt. Lett. **23**, 1701 (1998).
9. S. Longhi et al., Phys. Rev. E **67**, 036601 (2003).
10. K. Staliunas et al., Optics Express **14**, 10669 (2006).
11. K. Staliunas et al., Phys. Rev. E **73**, 016601 (2006).
12. H. S. Eisenberg et al., Phys. Rev. Lett. **81**, 3383 (1998).
13. R. Morandotti et al., Phys. Rev. Lett. **86**, 3296 (2001).
14. D. N. Christodoulides et al., Nature **424**, 817 (2003).
15. A. Szameit et al., Opt. Express **14**, 6055 (2006).
16. S. Longhi et al., Phys. Rev. Lett. **96**, 243901 (2006).
17. R. Iyer et al., Opt. Express **15**, 3212 (2007).
18. F. Dreisow et al., Opt. Express **16**, 3474 (2008).
19. I. L. Garanovich et al., Opt. Express **15**, 9737 (2007).
20. S. Longhi, Phys. Rev. A **71**, 065801 (2005).
21. G. della Valle et al., Phys. Rev. Lett. **98**, 263601 (2007).
22. X. Luo et al., Phys. Rev. A **76**, 051802(R) (2007).
23. I. L. Garanovich et al., Phys. Rev. Lett **100**, 203904 (2008).
24. S. Longhi and K. Staliunas, Opt. Commun. **281**, 4343 (2008).
25. A. Szameit et al., Appl. Phys. Lett. **90**, 241113 (2007).
26. S. Jensen, IEEE J. Quant. Electron. QE-18, 1580 (1982).
27. A. Szameit et al., Phys. Rev. A **78**, 031801 (2008).

# Figure captions

Figure 1. Experimental [(a),(c)] and theoretical [(b),(d)] intensity distributions for low-power excitation of the dual-core coupler. The upper edge of each panel corresponds to the input facet, the lower edge to the output facet. Panels (a) and (b) show unmodulated coupler, while panels (c) and (d) correspond to a modulated coupler with $\Omega/\Omega_{\rm b} = 1.25$ and $\mu = 0.2$.

Figure 2. (a) $\Omega_r/\Omega_b$ versus $\mu$ for a linear coupler. (b) $U_m$ versus $\Omega/\Omega_b$ in a linear coupler at $\mu = 0.2$. (c) $U_m$ versus $A^2$ in a coupler at $\mu = 0.1$ and $\Omega/\Omega_b = 0.76$. (d) $U_m$ in the surface channel (curve 1) or in the bulk (curve 2) of a linear array versus $\Omega/\Omega_b$ at $\mu = 0.2$.

Figure 3. Experimental (top) and theoretical (bottom) intensity distributions in an array for low-power excitation of surface waveguide [columns (a),(b)] and center waveguide [columns (c),(d)]. Panels (a) and (c) show unmodulated arrays, panels (b) and (d) show modulated arrays ($\mu = 0.2$) with $\Omega/\Omega_{\rm b} = 1.3$ for surface waveguide excitation and $\Omega/\Omega_{\rm b} = 1.38$ for the excitation in array center.

Figure 4. Experimental and theoretical intensity distributions for surface (a),(b) and center (c),(d) waveguide excitations. Theoretical plots, showing the propagation dynamics inside the sample, are placed on top of photographs showing experimental output intensity distributions. Panels (a) and (c) correspond to the unmodulated arrays, while in (b) $\mu = 0.15$ and $\Omega/\Omega_{\rm b} = 1.05$, and in (d) $\mu = 0.15$ and $\Omega/\Omega_{\rm b} = 1.07$. The peak powers from top to bottom are (a) $0.16\,{\rm MW}$, $0.93\,{\rm MW}$, and $1.41\,{\rm MW}$, (b) $0.16\,{\rm MW}$, $0.88\,{\rm MW}$, and $1.60\,{\rm MW}$, (c) $0.16\,{\rm MW}$, $0.85\,{\rm MW}$, and $1.41\,{\rm MW}$, and (d) $0.16\,{\rm MW}$, $1.01\,{\rm MW}$, and $2.13\,{\rm MW}$.

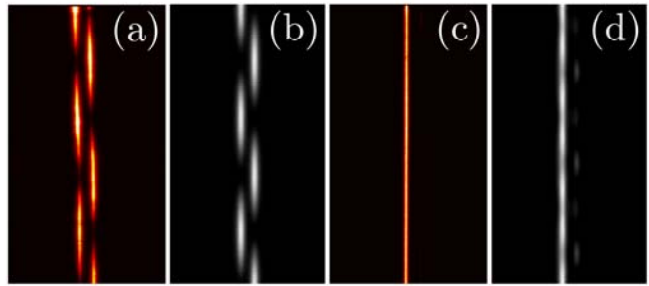

Figure 1. Experimental [(a),(c)] and theoretical [(b),(d)] intensity distributions for low-power excitation of the dual-core coupler. The upper edge of each panel corresponds to the input facet, the lower edge to the output facet. Panels (a) and (b) show unmodulated coupler, while panels (c) and (d) correspond to a modulated coupler with $\Omega/\Omega_b = 1.25$ and $\mu = 0.2$.

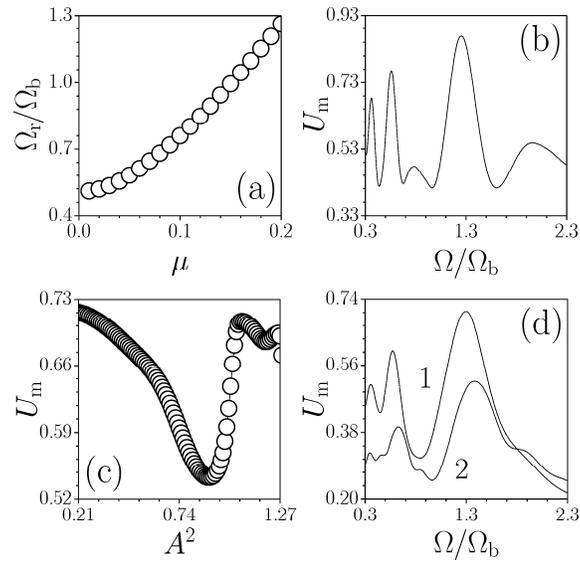

Figure 2. (a) $\Omega_r/\Omega_b$ versus $\mu$ for a linear coupler. (b) $U_m$ versus $\Omega/\Omega_b$ in a linear coupler at $\mu = 0.2$. (c) $U_m$ versus $A^2$ in a coupler at $\mu = 0.1$ and $\Omega/\Omega_b = 0.76$. (d) $U_m$ in the surface channel (curve 1) or in the bulk (curve 2) of a linear array versus $\Omega/\Omega_b$ at $\mu = 0.2$.

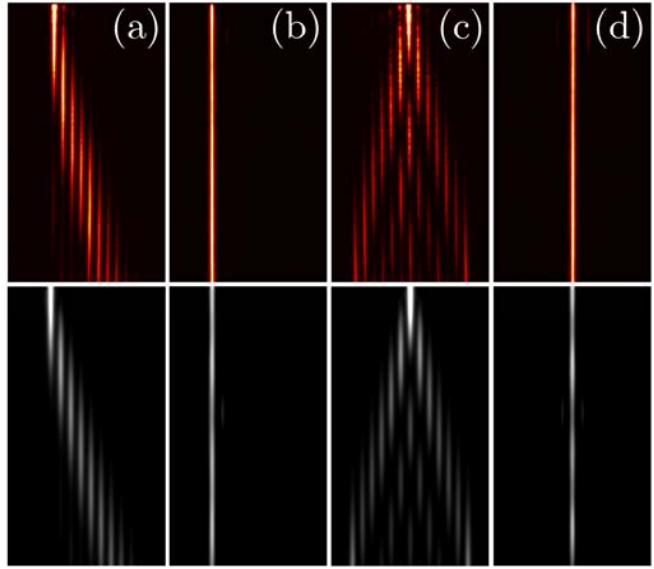

Figure 3. Experimental (top) and theoretical (bottom) intensity distributions in an array for low-power excitation of surface waveguide [columns (a),(b)] and center waveguide [columns (c),(d)]. Panels (a) and (c) show unmodulated arrays, panels (b) and (d) show modulated arrays ($\mu = 0.2$) with $\Omega/\Omega_b = 1.3$ for surface waveguide excitation and $\Omega/\Omega_b = 1.38$ for the excitation in array center.

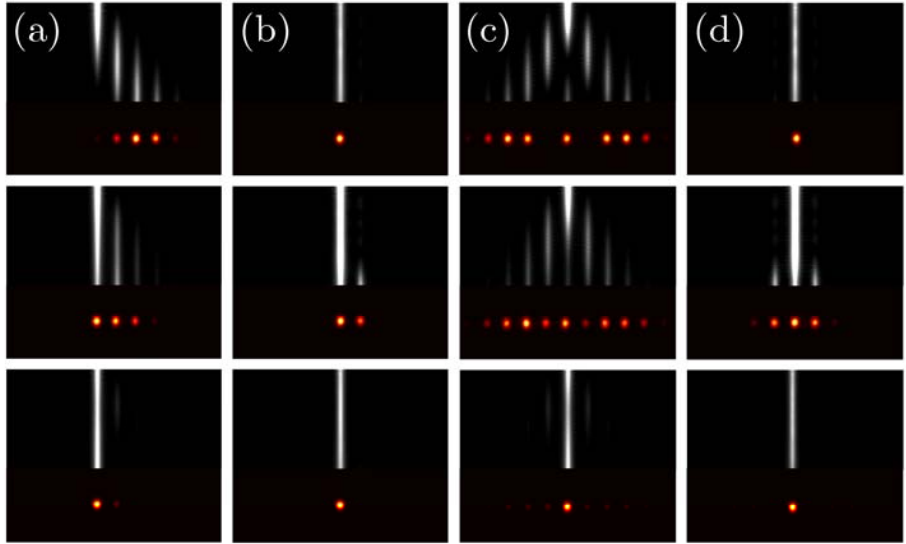

Figure 4. Experimental and theoretical intensity distributions for surface (a),(b) and center (c),(d) waveguide excitations. Theoretical plots, showing the propagation dynamics inside the sample, are placed on top of photographs showing experimental output intensity distributions. Panels (a) and (c) correspond to the unmodulated arrays, while in (b) $\mu = 0.15$ and $\Omega/\Omega_b = 1.05$, and in (d) $\mu = 0.15$ and $\Omega/\Omega_b = 1.07$. The peak powers from top to bottom are (a) $0.16\,\text{MW}$, $0.93\,\text{MW}$, and $1.41\,\text{MW}$, (b) $0.16\,\text{MW}$, $0.88\,\text{MW}$, and $1.60\,\text{MW}$, (c) $0.16\,\text{MW}$, $0.85\,\text{MW}$, and $1.41\,\text{MW}$, and (d) $0.16\,\text{MW}$, $1.01\,\text{MW}$, and $2.13\,\text{MW}$.